\documentclass{elsartNoFoot}

\usepackage{latin1}
\usepackage{amsmath}
\usepackage{amssymb}

\usepackage{jb}
\usepackage[british]{babel}
\usepackage{url}

\usepackage[pdftex]{color}
\usepackage{graphicx}		% \atsym

\usepackage[paperwidth=210mm,paperheight=297mm]{geometry}	% A4
\renewcommand{\arraystretch}{1.1}

\definecolor{coLabel}	{rgb}{0.99,0.50,0.99}	% label (debug only)
\definecolor{coRemark}	{rgb}{0.99,0.00,0.50}	% remark

\definecolor{coConstructor}	{rgb}{0.00,0.40,0.00}	% constructor name
\definecolor{coDefdFct}		{rgb}{0.00,0.00,0.40}	% function name
\definecolor{coSort}		{rgb}{0.40,0.00,0.00}	% sort name
\definecolor{coVar}		{rgb}{0.40,0.40,0.00}	% variable
\definecolor{coEqnLab}		{rgb}{0.99,0.00,0.00}	% equation label
\definecolor{coEqnRef}		{rgb}{0.60,0.20,0.20}	% equation ref

\newcommand{\LABEL}[1]{\label{#1}}

\newcommand{\raa}{\longrightarrow}

\renewcommand{\leq}{\leqslant}
\renewcommand{\geq}{\geqslant}

\newlog{\vars}{vars}

\newcommand{\abs}[1]{\left| #1 \right|}		% absolute value
\newcommand{\set}[1]{\{ #1 \}}			% set
\newcommand{\subst}[1]{\{ #1 \}}		% substitution

\renewcommand{\:}[4]{%
        {%
        \renewcommand{\:}[4]{%
                {%
                \renewcommand{\:}[4]{error\error}%
                \renewcommand{\j}{{##2}}%
                {##4}%
                ##1...##1%
                \renewcommand{\j}{{##3}}%
                {##4}%
                }%
        }%
        \renewcommand{\i}{{#2}}%
        {#4}%
        #1\ldots#1%
        \renewcommand{\i}{{#3}}%
        {#4}%
        }%
}

\newcommand{\notion}[1]{\emph{#1}}

\newcommand{\cn}[1]{\textcolor{coConstructor}{\sf #1}}	% constructor name
\newcommand{\sn}[1]{\textcolor{coSort}{\sf #1}}		% sort name
\newcommand{\fn}[1]{\textcolor{coDefdFct}{\sf #1}}	% function name
\newcommand{\vn}[1]{\textcolor{coVar}{\sf #1}}		% variable name
\newcommand{\sortdef}{::=}
\newcommand{\alt}{\mid}

\newcommand{\nat}{\sn{nat}}
\newcommand{\s}{\cn{s}}
\newcommand{\0}{\cn{0}}

\newcommand{\lst}{\sn{list}}
\newcommand{\nil}{\cn{nil}}
\newcommand{\cons}{\cn{::}}

\newcommand{\tree}{\sn{tree}}
\newcommand{\nul}{\cn{null}}
\newcommand{\node}{\cn{nd}}

\newcommand{\Blist}{\sn{blist}}
\newcommand{\Bnil}{\cn{nl}}
\newcommand{\Bo}{\cn{o}}
\newcommand{\Bi}{\cn{i}}

\newcommand{\add}{\fn{+}}
\newcommand{\mul}{\fn{*}}
\newcommand{\sq}{\fn{sq}}
\newcommand{\lgth}{\fn{lgth}}
\newcommand{\sz}{\fn{size}}
\newcommand{\dup}{\fn{dup}}
\newcommand{\app}{\fn{app}}
\newcommand{\Badd}{\fn{add}}
\newcommand{\Binc}{\fn{inc}}

\newcommand{\w}{\vn{w}}
\newcommand{\x}{\vn{x}}
\newcommand{\y}{\vn{y}}
\newcommand{\z}{\vn{z}}
\renewcommand{\a}{\vn{a}}
\renewcommand{\b}{\vn{b}}
\renewcommand{\c}{\vn{c}}

\newcounter{eqn}

\newcommand{\EQNSPACE}{\hspace*{1cm}}

\newcommand{\EQN}{%
	\refstepcounter{eqn}%
	\textcolor{coEqnLab}{\scriptstyle\mathbf\theeqn:\EQNSPACE}%
}

\newcommand{\REF}[1]{\textcolor{coEqnRef}{\mbox{\scriptsize\bf\ref{#1}}}}

\begin{document}

\newsavebox{\atpic}
\savebox{\atpic}{\includegraphics[scale=0.08]{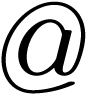}}

\setReceivedPrefix{}

\begin{frontmatter}

\title{A Scheme-Driven Approach to Learning Programs from Input/Output
Equations}
\author{Jochen Burghardt}
\address{jochen.burghardt\usebox{\atpic}alumni.tu-berlin.de}
\received{Feb 2018}

\begin{abstract}
We describe an approach to learn, in a term-rewriting setting, 
function definitions from input/output equations.
~
By confining ourselves to structurally recursive definitions we obtain
a fairly fast learning algorithm that often yields definitions close to
intuitive expectations.
~
We 
provide a {\sc Prolog} prototype implementation of our approach,
and indicate open issues of further investigation.
\end{abstract}


\end{frontmatter}

\vfill

\setcounter{tocdepth}{1}

{
\renewcommand{\baselinestretch}{0.0}
\tableofcontents
}

\vfill

\newpage
\section{Introduction}
\LABEL{Introduction}

This paper describes an approach to learn function definitions
from input/output equations.\footnote{
	We will use henceforth ``i/o equations'' for brevity.
	~
	We avoid calling them ``examples'' as
	this could cause confusion when we explain
	our approach along example sort definitions, example 
	signatures, and example functions.
}
~
In trivial cases, a definition is obtained by syntactical
anti-unification of the given i/o equations.
~
In non-trivial cases, we assume a structurally recursive
function
definition, and transform the given i/o equations into equations for the
employed auxiliary functions.
~
The latter are learned from their i/o equations in turn, until a
trivial case is reached.

We came up with this
approach in 1994 but didn't publish it until today.
~
In this paper, we explain it mainly along some learning examples,
leaving a theoretical elaboration to be done.
~
Also, we indicate several issues of improvement
that should be investigated further.
~
However, we provide at least a {\sc Prolog} 
prototype implementation of our approach.

In the rest of this section, we introduce the term-rewriting setting
our approach works in.
~
In Sect.~\ref{The task of learning functions},
we define the task of function learning.
~
In Sect.~\ref{Learning functions by anti-unification}
and~\ref{Learning functions by structural recursion},
we explain the base case and the inductive case of our approach,
that is,
how to learn trivial functions, and how to reduce learning
sophisticated functions to learning easier functions,
respectively.
~
Section~\ref{Possible extensions} sketches some ideas for possible
extensions to our approach; it also shows its limitations.
~
Some runs of our {\sc Prolog} prototype are shown in
Appendix~\ref{Example runs of our prototype implementation}.

\begin{figure}[hb]
\begin{center}
$$\begin{array}{r| rcccccc |l}
\cline{2-8}
\EQN\LABEL{sort nat}
	& \nat & \sortdef & \0 & \alt & \s(\nat)
	&&& \mbox{ natural numbers}	\\
\EQN\LABEL{sort list}
	& \lst & \sortdef & \nil & \alt & \nat \cons \lst
	&&& \mbox{ lists of natural numbers}	\\
\EQN\LABEL{sort tree}
	& \tree & \sortdef & \nul & \alt & \node(\tree,\nat,\tree)
	&&& \mbox{ binary trees of natural numbers}	\\
\EQN\LABEL{sort blist}
	& \Blist & \sortdef & \Bnil & \alt & \Bo(\Blist) & \alt &
	\Bi(\Blist)
	& \mbox{ list of binary digits}	\\
\cline{2-8}
\end{array}$$
\caption{Employed sort definitions}
\LABEL{Employed sort definitions}
\end{center}
\end{figure}

We use a term-rewriting setting 
that is well-known from functional programming:
~
A \notion{sort} can be defined recursively by giving its
\notion{constructors}.
~
For example, sort definition~\REF{sort nat}, shown in
Fig.~\ref{Employed sort definitions},
defines the sort $\nat$ of all natural numbers in $\0$-$\s$ notation.
~
In this example, we use $\0$ as a nullary, and $\s$ as a unary constructor.

\begin{figure}[hb]
\begin{center}
$$\begin{array}{r| rcccc |l}
\cline{2-6}
\EQN\LABEL{sgn add} & \add & : & \nat \times \nat & \raa & \nat
	& \mbox{ addition of natural numbers}	\\
\EQN\LABEL{sgn mul} & \mul & : & \nat \times \nat & \raa & \nat
	& \mbox{ multiplication of natural numbers}	\\
\EQN\LABEL{sgn lgth} & \lgth & : & \lst & \raa & \nat
	& \mbox{ number of elements of a list}	\\
\EQN\LABEL{sgn app} & \app & : & \lst \times \lst & \raa & \lst
	& \mbox{ concatenation of lists}	\\
\EQN\LABEL{sgn size} & \sz & : & \tree & \raa & \nat
	& \mbox{ number of elements of a binary tree}	\\
\EQN\LABEL{sgn dup} & \dup & : & \nat & \raa & \nat
	& \mbox{ duplicating a natural number}	\\
\EQN\LABEL{sgn badd}
	& \Badd & : & \Blist \times \Blist & \raa & \Blist
	& \mbox{ addition of binary numbers (lists)}	\\
\cline{2-6}
\end{array}$$
\caption{Employed function signatures}
\LABEL{Employed function signatures}
\end{center}
\end{figure}

A sort is understood as representing a possibly infinite
set of ground constructor terms,\footnote{
	i.e.\ terms without \notion{variables}, 
	built only from constructor symbols
}
e.g.\
the sort $\nat$ represents the set
$\set{ \0, \s(\0), \s(\s(\0)), \s(\s(\s(\0))), \ldots }$.
~
A \notion{function} has a fixed \notion{signature};
Fig.~\ref{Employed function signatures} gives some examples.
~
The signature of a constructor can be inferred from the sort
definition it occurs in, e.g.\
$\0: \raa \nat$ and $\s: \nat \raa \nat$.
~
We don't allow non-trivial equations between constructor terms,
therefore, we have $T_1 = T_2$ iff $T_1$ syntactically equals $T_2$,
for all ground constructor terms $T_1$, $T_2$.

\begin{figure}
\begin{center}
$$\begin{array}{r |rcl|}
\cline{2-4}
\EQN\LABEL{add def 0}
	& \x \add \0 & = & \x	\\
\EQN\LABEL{add def s}
	& \x \add \s(\y) & = & \s(\x \add \y)	\\
\EQN\LABEL{mul def 0}
	& \x \mul \0 & = & \0	\\
\EQN\LABEL{mul def s}
	& \x \mul \s(\y) & = & \x \add \x \mul y	\\
\EQN & \lgth(\nil) & = & \0	\\
\EQN & \lgth(\x \cons \y) & = & \s(\lgth(\y))	\\
\EQN\LABEL{app def nil}
	& \app(\nil,\z) & = & \z	\\
\EQN\LABEL{app def cons}
	& \app(\x \cons \y, \z) & = & \x \cons \app(\y,\z)	\\
\EQN & \sz(\nul) & = & \0	\\
\EQN & \sz(\node(\x,\y,\z)) & = & \s(\sz(\x) \add \sz(\z))	\\
\cline{2-4}
\end{array}$$
\caption{Example function definitions}
\LABEL{Example function definitions}
\end{center}
\end{figure}

A non-constructor function
can be \notion{defined} by giving a terminating
(\cite[Sect.5.1, p.270]{Dershowitz.Jouannaud.1990a})
\notion{term rewriting system} for it
such that its left-hand sides are sufficiently complete
(\cite{Guttag.1977}, \cite{Comon.1986}, 
\cite[Sect.3.2, p.264]{Dershowitz.Jouannaud.1990a}).
~
Examples for function definitions are
shown in Fig.~\ref{Example function definitions}.
~
Given some functions $\:,1m{f_\i}$ defined by such a term rewriting system,
for each $i$ and each ground constructor terms $\:,1n{T_\i}$ we can
find a unique ground constructor term $T$
such that $f_i(\:,1n{T_\i}) = T$.
~
We then say that $f_i(\:,1n{T_\i})$ \notion{evaluates} to $T$.

Given a term $T$, we denote by $\vars(T)$ the set of variables
occurring in $T$.

\section{The task of learning functions}
\LABEL{The task of learning functions}

The problem our approach shall solve is the following.
~
Given a set of sort definitions, a non-constructor function symbol $f$,
its signature, and a set of
input/output equations

for $f$, construct a term rewriting
system defining $f$ such that it behaves as prescribed by the
i/o equations.
~
We say that we want to \notion{learn} a definition for $f$,
or sloppily, that we want to learn $f$, from the given i/o equations.

For example, given sort definition \REF{sort nat}, 
signature \REF{sgn dup}, and the following input/output ground equations
$$\begin{array}{rrcl}
\EQN\LABEL{exm dup 0}
	& \dup( \0 ) & = & \0	\\
\EQN \LABEL{exm dup 1}
	& \dup( \s(\0) ) & = & \s(\s(\0))	\\
\EQN \LABEL{exm dup 2}
	& \dup( \s(\s(\0)) ) & = & \s(\s(\s(\s(\0))))	\\
\EQN \LABEL{exm dup 3}
	& \dup( \s(\s(\s(\0))) ) & = & \s(\s(\s(\s(\s(\s(\0))))))	\\
\end{array}$$
we are looking for a definition of $\dup$ such that
equations \REF{exm dup 0}, \REF{exm dup 1}, \REF{exm dup 2},
and~\REF{exm dup 3} hold.
~
One such definition is
$$\begin{array}{rrcl}
\EQN & \dup( \0 ) & = & \0   \\
\EQN & \dup( \s(\x) ) & = & \s(\s(\dup(\x)))	\\
\end{array}$$
We say that this definition \notion{covers} the i/o equations 
\REF{exm dup 0}, \REF{exm dup 1}, \REF{exm dup 2}, and~\REF{exm dup 3}.
~
In contrast, a definition
$$\begin{array}{rrcl}
\EQN & \dup( \0 ) & = & \0   \\
\EQN & \dup( \s(\x) ) & = & \s(\s(\x))	\\
\end{array}$$
would cover i/o equations \REF{exm dup 0} and \REF{exm dup 1},
but neither~\REF{exm dup 2} nor~\REF{exm dup 3}.
~
We wouldn't accept this definition, since 
we are interested only in function definitions that cover {\em all}
given i/o equations.

It is well-known that
%---~often contrary to superficial human intuition~---
there isn't a unique solution to our problem.
~
In fact, given i/o equations
$\:,1n{f(L_\i) = R_\i}$ and an arbitrary
function $g$ of appropriate domain and range,
e.g.\ the function defined by\footnote{
	We use common imperative notation here for sake of
	readability.
}
$$f(x) = 
(
\mbox{ \sf if } x = L_1 \mbox{ \sf then } R_1
\mbox{ \sf elif }
\ldots
\mbox{ \sf elif } x = L_n \mbox{ \sf then } R_n
\mbox{ \sf else } g(x) \mbox{ \sf fi }
)$$
trivially covers all i/o equations.
~
%In the setting of learning predicate definitions from examples,
%Watanabe \cite[\REM{ugly duckling theprem}]{Watanabe.1969}
%argued that preferring one definition to another can at be be
%justified by extra-logical reasons.
%~
Usually, the ``simplest'' function definitions are preferred, with
``simplicity'' being some user-defined measure loosely corresponding
to term size and/or case-distinction count,
like e.g.\ in \cite[p.8]{Burghardt.2005c} and
\cite[p.77]{Kitzelmann.2010}.
~
However, the notion of simplicity depends on the language of available
basic operations.\footnote{
	The ``invariance theorem'' in  Kolmogorov complexity theory
	(e.g.\ \cite[p.105, Thm.2.1.1]{Li.Vitanyi.2008})
	implies that 
	$\forall L_1, L_2 \; \exists c \; \forall x: \;
	\abs{ C_{L_1}(x) - C_{L_2}(x)} \leq c$,
	where the $L_i$ range over Turing-complete algorithm description
	languages,
	$c$ is a natural number,
	$x$ ranges over i/o equation sets,
	and $C_L(x)$ denotes the length of the shortest function
	definition,
	written in $L$, that covers $x$.
	~
	This theorem
	is sometimes misunderstood to enable a language-independent
	notion of simplicity; however, it
	does not, at least for small i/o example sets.
}
~
%Again in a philosophical predicate-learning setting,
%Goodman has argued in favour of this dependency
%in his famous ``grue emeralds'' example.
%\cite[\REM{page}]{Goodman.1973}, \cite[\REM{page}]{Goodman.1946}
In the end, the notion of a ``good'' definition can hardly be defined
more precisely than being one that meets common human prejudice.
~
From our prototype runs
we got the feeling that our approach often yields ``good''
definition in that sense.
%~
%Proper psychological
%experiments should be done to confirm or disconfirm that
%feeling, based on a larger set of learned functions, and a larger set
%of humans assessing their definitions.

\section{Learning functions by anti-unification}
\LABEL{Learning functions by anti-unification}

One of the simplest ways to obtain a function definition is to
syntactically anti-unify the given i/o equations.
$$\begin{array}{l @{\;\;\;\;\;} c@{}c@{}c@{}c@{}c c c l}
\mbox{Given i/o equations}
	& f( & L_{11} & , \ldots , & L_{m1} & ) & = & R_1	\\[-1ex]
	&& \vdots && \vdots &&& \vdots	\\[-1ex]
	& f( & L_{1n} & , \ldots , & L_{mn} & ) & = & R_n & ,	\\
\cline{2-8}
\\[-2ex]
\mbox{let}
	& f( & L_1 & , \ldots , & L_m & ) & = & R	\\
\end{array}$$
be their \notion{least general generalization} 
(\notion{lgg} for short,
see \cite{Plotkin.1970a,Plotkin.1971,Reynolds.1970}).
~
If the variable condition $\vars(R) \subseteq \:{\cup}1m{ \vars(L_\i) }$
holds,
then the lgg will cover all $n$ given i/o equations.

For example, assume we are to generate a definition for a unary function
called $\fn{g_2}$
$$\begin{array}{l @{\;\;\;\;\;} r c@{}c@{}c c c@{}c@{}c c}
\mbox{from the i/o equations}
& \EQN\LABEL{exm g2 0}
	& \fn{g_2}( & \0 & ) & = & \s(\s( & \0 & ))	\\
& \EQN\LABEL{exm g2 2}
	& \fn{g_2}( & \s(\s(\0)) & ) & = & \s(\s( & \s(\s(\0)) & ))	\\
& \EQN\LABEL{exm g2 4}
	& \fn{g_2}( & \s(\s(\s(\s(\0)))) & ) & = 
	& \s(\s( & \s(\s(\s(\s(\0)))) & )) & .	\\
\cline{3-9}
\\[-2ex]
\mbox{We obtain the lgg}
& \EQN\LABEL{g2 def}
	& \fn{g_2}(&\x_{024}&) & = & \s(\s( & \x_{024} & ))	\\
\end{array}$$

As another example,
we can generate a definition for a binary function called $\fn{g_4}$
$$\begin{array}{l @{\;\;\;\;\;} r c@{}c@{}c@{}c@{}c c c@{}c@{}c c}
\mbox{from the i/o equations}
& \EQN\LABEL{exm g4 0}
	& \fn{g_4}(&\a&,&\0&) & = & \s(&\0&) 	\\
& \EQN\LABEL{exm g4 1}
	& \fn{g_4}(&\a&,&\s(\0)&) & = & \s(&\s(\0)&)	\\
& \EQN\LABEL{exm g4 2}
	& \fn{g_4}(&\a&,&\s(\s(\0))&) & = & \s(&\s(\s(\0))&) & .	\\
\cline{3-11}
\\[-2ex]
\mbox{We obtain the lgg}
& \EQN\LABEL{g4 def}
	& \fn{g_4}(&\a&,&\y_{012}&) & = & \s(&\y_{012}&)	\\
\end{array}$$
which satisfies the variable condition.\footnote{
	Whenever applied to terms $\:,1m{T_\i}$ that don't start all
	with the same function symbol,
	Plotkin's \notion{lgg} algorithm returns a variable
	that uniquely depends on $\:,1m{T_\i}$.
	~
	We indicate the originating terms by an index sequence;
	e.g.\ $\y_{012}$ was obtained as 
	${\it lgg}(\0,\s(\0),\s(\s(\0)))$.
}
~
Hence when $\fn{g_4}$ is defined by equation~\REF{g4 def},
it covers i/o equations~\REF{exm g4 2}, \REF{exm g4 1},
and~\REF{exm g4 0}.

As a counter-example, the lgg of the above
$$\begin{array}{l @{\;\;\;\;\;} r c@{}c@{}c c c}
\mbox{$\dup$ i/o equations}
& \REF{exm dup 0}\EQNSPACE
	& \dup( & \0 & ) & = & \0	\\
& \REF{exm dup 1}\EQNSPACE
	& \dup( & \s(\0) & ) & = & \s(\s(\0))	\\
& \REF{exm dup 2}\EQNSPACE
	& \dup( & \s(\s(\0)) & ) & = & \s(\s(\s(\s(\0))))	\\
& \REF{exm dup 3}\EQNSPACE
	& \dup( & \s(\s(\s(\0))) & ) & = & \s(\s(\s(\s(\s(\s(\0))))))	\\
\cline{3-7}
\\[-2ex]
\mbox{is computed as}
& \EQN & \dup( & \x_{0123} & ) & = & \x_{0246}	\\
\end{array}$$
which violates the variable condition, and thus cannot be used to
reduce a term $\dup(T)$ to a ground constructor term, i.e.\ to
evaluate $\dup(T)$.

The above anti-unification approach can be extended in several
ways, they are sketched in Sect.~\ref{Extension of anti-unification}.
~
However, in all but trivial cases, an lgg will violate the
variable condition, and we need another approach to learn a function
definition.

\section{Learning functions by structural recursion}
\LABEL{Learning functions by structural recursion}

For a function $f$ that can't be learned by
Sect.~\ref{Learning functions by anti-unification},
we assume a defining term rewriting system that follows a
\notion{structural recursion} scheme obtained from $f$'s signature and
a guessed argument position.

For example, for the function $\dup$ with the signature given
in~\REF{sgn dup} and the only possible argument position, 1, we obtain
the schematic equations
$$\begin{array}{rrcl}
\EQN\LABEL{rec dup 1} & \dup(\0) & = & \fn{g_1}	\\
\EQN\LABEL{rec dup 2} & \dup(\s(\x)) & = & \fn{g_2}(\dup(\x))	\\
\end{array}$$
where $\fn{g_1}$ and $\fn{g_2}$ are fresh names of non-constructor
functions.

If we could learn appropriate definitions for $\fn{g_1}$ and
$\fn{g_2}$,
we could obtain a definition for $\dup$ 
just by adding equations~\REF{rec dup 1} and~\REF{rec dup 2}.
~
The choice of $\fn{g_1}$ is obvious:
$$\begin{array}{r cccccc}
\EQN\LABEL{g1 def}
	\0
	& \stackrel{\REF{exm dup 0}}{=} & \dup(\0)
	& \stackrel{\REF{rec dup 1}}{=} & \fn{g_1}	\\
\end{array}$$
In order to learn a definition for $\fn{g_2}$, we need to obtain 
appropriate i/o examples for $\fn{g_2}$ from those for $\dup$.
~
Joining equation~\REF{rec dup 2} with $\dup$'s relevant i/o equations
yields three i/o equations for $\fn{g_2}$:
$$\begin{array}{r ccccccc}
\REF{exm g2 0}\EQNSPACE
	& \s(\s(\0)) 
	& \stackrel{\REF{exm dup 1}}{=} & \dup( \s(\0) )
	& \stackrel{\REF{rec dup 2}}{=} & \fn{g_2}(\dup(\0))
	& \stackrel{\REF{exm dup 0}}{=} & \fn{g_2}(\0)	\\
\REF{exm g2 2}\EQNSPACE
	& \s(\s(\s(\s(\0)))) 
	& \stackrel{\REF{exm dup 2}}{=} & \dup( \s(\s(\0)) )
	& \stackrel{\REF{rec dup 2}}{=} & \fn{g_2}(\dup(\s(\0)))
	& \stackrel{\REF{exm dup 1}}{=} & \fn{g_2}(\s(\s(\0)))	\\
\REF{exm g2 4}\EQNSPACE
	& \s(\s(\s(\s(\s(\s(\0)))))) 
	& \stackrel{\REF{exm dup 3}}{=} & \dup( \s(\s(\s(\0))) )
	& \stackrel{\REF{rec dup 2}}{=} & \fn{g_2}(\dup(\s(\s(\0))))
	& \stackrel{\REF{exm dup 2}}{=} & \fn{g_2}(\s(\s(\s(\s(\0))))	\\
\end{array}$$
A definition for $\fn{g_2}$ covering its i/o examples~\REF{exm g2 0},
\REF{exm g2 2}, and~\REF{exm g2 4} has already been derived by
anti-unification in
Sect.~\ref{Learning functions by anti-unification} as
$$\begin{array}{r ccc l}
\REF{g2 def}\EQNSPACE
	& \fn{g_2}(\x_{024}) & = & \s(\s(\x_{024})) & .	\\
\end{array}$$
Altogether, we obtain the rewriting system
$$\begin{array}{r ccc l}
\REF{rec dup 1}\EQNSPACE & \dup(\0) & = & \fn{g_1}	\\
\REF{rec dup 2}\EQNSPACE & \dup(\s(\x)) & = & \fn{g_2}(\dup(\x))	\\
\REF{g1 def}\EQNSPACE & \fn{g_1} & = & \0	\\
\REF{g2 def}\EQNSPACE & \fn{g_2}(\x_{024}) & = & \s(\s(\x_{024}))	\\
\end{array}$$
as a definition for $\dup$ that covers its i/o 
equations~\REF{exm dup 0},
\REF{exm dup 1}, 
\REF{exm dup 2}, and~\REF{exm dup 3}. 
~
Subsequently, this system may be simplified, by \notion{inlining}, to
$$\begin{array}{r ccc l}
\REF{rec dup 1}\EQNSPACE & \dup(\0) & = & \0	\\
\REF{rec dup 2}\EQNSPACE & \dup(\s(\x)) & = & \s(\s(\dup(\x)))	\\
\end{array}$$
which is the usual definition of the $\dup$ function.

Returning to the computation of i/o equations for $\fn{g_2}$ from
those for $\dup$,
note that $\fn{g_2}$'s derived i/o 
equations~\REF{exm g2 0}, \REF{exm g2 2}, and~\REF{exm g2 4}
were \notion{necessary} in the sense that they must be satisfied by
each possible definition
of $\fn{g_2}$ that leads to $\dup$ covering its i/o 
equations (\REF{exm dup 1}, \REF{exm dup 2}, and~\REF{exm dup 3}).
~
Conversely, $\fn{g_2}$'s i/o equations were also \notion{sufficient}
in the sense that each possible definition of $\fn{g_2}$ covering them
ensures that $\dup$ covers~\REF{exm dup 1}, \REF{exm dup 2},
and~\REF{exm dup 3},
provided it covers~\REF{exm dup 0}:
$$\begin{array}{r ccccccc}
\mbox{Proof of~\REF{exm dup 1}:\EQNSPACE}
	& \dup( \s(\0) )
	& \stackrel{\REF{rec dup 2}}{=} & \fn{g_2}(\dup(\0))
	& \stackrel{\REF{exm dup 0}}{=} & \fn{g_2}(\0)
	& \stackrel{\REF{exm g2 0}}{=} & \s(\s(\0))	\\
\mbox{Proof of~\REF{exm dup 2}:\EQNSPACE}
	& \dup( \s(\s(\0)) )
	& \stackrel{\REF{rec dup 2}}{=} & \fn{g_2}(\dup(\s(\0)))
	& \stackrel{\REF{exm dup 1}}{=} & \fn{g_2}(\s(\s(\0)))
	& \stackrel{\REF{exm g2 2}}{=} & \s(\s(\s(\s(\0))))	\\
\mbox{Proof of~\REF{exm dup 3}:\EQNSPACE}
	& \dup( \s(\s(\s(\0))) )
	& \stackrel{\REF{rec dup 2}}{=} & \fn{g_2}(\dup(\s(\s(\0))))
	& \stackrel{\REF{exm dup 2}}{=} & \fn{g_2}(\s(\s(\s(\s(\0)))))
	& \stackrel{\REF{exm g2 4}}{=} & \s(\s(\s(\s(\s(\s(\0))))))	\\

\end{array}$$
Observe that the above proofs are based just on permutations of the
equation chains from~\REF{exm g2 0}, \REF{exm g2 2}, and~\REF{exm g2 4}.
~
Moreover, note that the coverage proof for $\dup(\s(T))$ relies on 
the coverage for $\dup(T)$ already being proven.
~
That is, the coverage proofs follow the employed
structural recursion scheme.
~
As for the base case, $\fn{g_1}$'s coverage of~\REF{g1 def} is
of course
necessary and sufficient for $\dup$'s coverage of~\REF{exm dup 0}.

\subsection{Non-ground i/o equations}

As an example that uses i/o equations containing variables,
consider the function $\lgth$, with the signature given
in~\REF{sgn lgth}.
~
Usually, i/o equations for this functions are given in a way that
indicates that the particular values of the list elements don't
matter.
~
For example, an i/o equation like 
$\lgth( \a \cons \b \cons \nil) = \s(\s(\0))$
is seen much more often than
$\lgth( \s(\0) \cons \0 \cons \nil) = \s(\s(\0))$.
~
Our approach allows for variables in i/o equations, and treats them as
universally quantified.
~
That is, a non-ground i/o equation is covered by a function definition
iff all its ground instances are.

Assume for example we are given the i/o equations
$$\begin{array}{rrcll}
\EQN\LABEL{exm lgth nil} & \lgth( \nil) & = & \0 & .	\\
\EQN\LABEL{exm lgth a} & \lgth( \a \cons \nil) & = & \s(\0)	\\
\EQN\LABEL{exm lgth ab}
	& \lgth( \a \cons \b \cons \nil) & = & \s(\s(\0))	\\
\EQN\LABEL{exm lgth abc}
	& \lgth( \a \cons \b \cons \c \cons \nil) & = & \s(\s(\s(\0)))	\\
\end{array}$$
Given the signature of $\lgth$
(see~\REF{sgn lgth}) and argument position 1, we obtain a structural
recursion scheme
$$\begin{array}{rrcll}
\EQN\LABEL{rec lgth 1} & \lgth(\nil) & = & \fn{g_3}	\\
\EQN\LABEL{rec lgth 2}
	& \lgth(\x \cons \y) & = & \fn{g_4}(\x,\lgth(\y)) & .	\\
\end{array}$$
Similar to the $\dup$ example, we get
$$\begin{array}{r cccccc}
\EQN\LABEL{g3 def}
	\0
	& \stackrel{\REF{exm lgth nil}}{=} & \lgth(\nil)
	& \stackrel{\REF{rec lgth 1}}{=} & \fn{g_3} & ,	\\
\end{array}$$
and
we can obtain i/o equations for $\fn{g_4}$ from those for
$\lgth$:\footnote{
	In the rightmost equation of each line, we employ a renaming
	substitution.
	~
	For example,
	we apply $\subst{\a \mapsto \b, \b \mapsto \c}$
	to i/o equation~\REF{exm lgth ab}
	in line~\REF{exm g4 2}.
	~
	For this reason, our approach wouldn't work if $\a$, $\b$, $\c$
	were considered non-constructor constants rather than
	universally quantified variables.
}
$$\begin{array}{cccccccc}
\REF{exm g4 0}\EQNSPACE
	& \s(\0)
	& \stackrel{\REF{exm lgth a}}{=}
	& \lgth( \a \cons \nil)
	& \stackrel{\REF{rec lgth 2}}{=} 
	& \fn{g_4}(\a,\lgth(\nil)) 
	& \stackrel{\REF{exm lgth nil}}{=} 
	& \fn{g_4}(\a,\0) \\
\REF{exm g4 1}\EQNSPACE
	& \s(\s(\0))
	& \stackrel{\REF{exm lgth ab}}{=}
	& \lgth( \a \cons \b \cons \nil)
	& \stackrel{\REF{rec lgth 2}}{=} 
	& \fn{g_4}(\a,\lgth(\b \cons \nil)) 
	& \stackrel{\REF{exm lgth a}}{=} 
	& \fn{g_4}(\a,\s(\0)) \\
\REF{exm g4 2}\EQNSPACE
	& \s(\s(\s(\0)))
	& \stackrel{\REF{exm lgth abc}}{=}
	& \lgth( \a \cons \b \cons \c \cons \nil)
	& \stackrel{\REF{rec lgth 2}}{=} 
	& \fn{g_4}(\a,\lgth(\b \cons \c \cons \nil)) 
	& \stackrel{\REF{exm lgth ab}}{=} 
	& \fn{g_4}(\a,\s(\s(\0))) \\

\end{array}$$
Again, a function definition covering these i/o equation happens to
have been derived by anti-unification in 
Sect.~\ref{Learning functions by anti-unification}:
$$\begin{array}{r ccc}
\REF{g4 def}\EQNSPACE
        & \fn{g_4}(\a,\y_{012}) & = & \s(\y_{012})  \\
\end{array}$$
Altogether, equations~\REF{rec lgth 1}, \REF{rec lgth 2},
\REF{g3 def}, and~\REF{g4 def}
build a rewriting system for $\lgth$ that covers all its given i/o
equations.
~
By subsequently inlining $\fn{g_3}$'s and
$\fn{g_4}$'s definition, we obtain a
simplified definition for $\lgth$:
$$\begin{array}{rrcl}
\EQN & \lgth(\nil) & = & \0      \\
\EQN & \lgth(\x \cons \y) & = & \s(\lgth(\y)) \\
\end{array}$$
which agrees with the usual one found in textbooks.

Similar to the ground case, $\fn{g_4}$'s derived i/o
equations~\REF{exm g4 0}, \REF{exm g4 1}, and~\REF{exm g4 2}
were necessary for $\lgth$ covering its i/o equations.
~
And as in the ground case, they are also sufficient:
$$\begin{array}{r ccccccc}
\mbox{Proof of~\REF{exm lgth a}:\EQNSPACE}
	& \lgth( \a \cons \nil )
	& \stackrel{\REF{rec lgth 2}}{=} & \fn{g_4}(\a,\lgth(\nil))
	& \stackrel{\REF{exm lgth nil}}{=} & \fn{g_4}(\a,\0)
	& \stackrel{\REF{exm g4 0}}{=} & \s(\0)     \\
\mbox{Proof of~\REF{exm lgth ab}:\EQNSPACE}
	& \lgth( \a \cons \b \cons \nil )
	& \stackrel{\REF{rec lgth 2}}{=}
	& \fn{g_4}(\a,\lgth(\b \cons \nil))
	& \stackrel{\REF{exm lgth a}}{=}
	& \fn{g_4}(\a,\s(\0))
	& \stackrel{\REF{exm g4 1}}{=} & \s(\s(\0))     \\
\mbox{Proof of~\REF{exm lgth abc}:\EQNSPACE}
	& \lgth( \a \cons \b \cons \c \cons \nil )
	& \stackrel{\REF{rec lgth 2}}{=}
	& \fn{g_4}(\a,\lgth(\b \cons \c \cons \nil))
	& \stackrel{\REF{exm lgth ab}}{=} 
	& \fn{g_4}(\a,\s(\s(\0))))
	& \stackrel{\REF{exm g4 2}}{=} 
	& \s(\s(\s(\0)))     \\
\end{array}$$
Again, renaming substitutions were used in the application 
of~\REF{exm lgth a} and~\REF{exm lgth ab}.

\subsection{Functions of higher arity}

For functions with more than one argument, we have several
choices of the argument on which to do the recursion.
~
In these cases, we currently systematically try all argument
positions\footnote{
	In particular, the recursive argument's sort and the
	function's result sort needn't be related in any way, as the
	$\lgth$ example above demonstrates.
}
in succession.
~
This is feasible since
\begin{itemize}
\item our approach is quite simple, and hence fast to compute, and
\item we have a sharp and easy to compute
	criterion (viz.\ coverage\footnote{
		Checking if an i/o equation is covered by a definition
		requires \notion{executing}
		the latter on the lhs arguments of
		the former.
		~
		Our structural recursion approach ensures the
		termination of such computations, and establishes an
		upper bound for the number of rewrite steps.
		~
		For example, $\fn{g_2}$ and $\fn{g_4}$, 
		defined in~\REF{g2 def} and~\REF{g4 def}, respectively,
		need one such step, while their callers $\dup$ and
		$\lgth$,
		defined in~\REF{rec dup 1},\REF{rec dup 2}
		and~\REF{rec lgth 1},\REF{rec lgth 2},
		respectively, need a linear amount of steps.
		~
		An upper-bound expression for learned functions' time
		complexity remains to be defined and proven.
	}
	of all i/o examples)
	to decide whether recursion on a given argument was successful.
\end{itemize}

For the function $\add$, with the signature given in~\REF{sgn add},
and argument position 2, we obtain the structural recursion scheme
$$\begin{array}{rrcll}
\EQN & \x \add \0 & = & \fn{g_5}(\x)	\\
\EQN & \x \add \s(\y) & = & \fn{g_6}(\x,\x \add \y) & .	\\
\end{array}$$
Appendix~\ref{Addition of $0$-$s$ numbers}
shows a run of our {\sc Prolog} prototype implementation
that obtains a definition for $\add$.
~
In Sect.~\ref{Extension of structural recursion}, we discuss possible
extensions of the structural recursion scheme, like
simultaneous recursion.

\subsection{Constructors with more than one recursion argument}

When computing a structural recursion scheme, we may encounter a sort
$s$ with a constructor that takes more than one argument of sort $s$.
~
A common example is the sort of all binary trees (of natural numbers),
as given in~\REF{sort tree}.
~
The function $\sz$, with the signature given in~\REF{sgn size},
computes the size of such a tree, i.e.\ the total number of $\node$
nodes.
~
A recursion scheme for the $\sz$ and argument position 1 looks like:
$$\begin{array}{rrcl}
\EQN & \sz(\nul) & = & \fn{g_9}	\\
\EQN & \sz(\node(\x,\y,\z)) & = & \fn{g_{10}}(\y,\sz(\x),\sz(\z))	\\
\end{array}$$

In App.~\ref{Size of a tree}, we show a prototype
run to obtain a definition for $\sz$.

\subsection{General approach}
\LABEL{General approach}

In the previous sections, we have introduced our approach using
particular examples.
~
In this section, we sketch a more abstract and algorithmic description.

Given a function and its signature
$f : \:{\times}1n{s_\i} \raa s$, 
and given one of its argument positions $1 \leq i \leq n$,
we can easily obtain a term rewriting system 
to define $f$ by \notion{structural recursion} on its $i$th argument.
~
Assume in the definition of $f$'s $i$th domain sort $s_i$ we
have an alternative
$$s_i \sortdef \ldots \mid c( \:,1l{ s'_\i } ) \mid \ldots ,$$
assume $\set{ \:,1m{s'_{\nu(\i)}} } \not\ni s_i$ is the set of
non-recursive arguments of the constructor $c$,
and $\:=1k{s'_{\rho(\i)}} = s_i$ are the recursive arguments of $c$.
~
Let $g$ be a new function symbol.
~
We build an equation
$$\begin{array}{r@{}l}
f( & \:,1{i-1}{x_\i} , c( \:,1l{y_\i} ) , \:,{i+1}n{x_\i} )	\\
=	\\
g( &
	\:,1{i-1}{x_\i} , \:,{i+1}n{x_\i} , \:,1m{y_{\nu(\i)}} ,	\\
& f( \:,1{i-1}{x_\i} , y_{\rho(1)} , \:,{i+1}n{x_\i} )	\\
& \ldots	\\
& f( \:,1{i-1}{x_\i} , y_{\rho(k)} , \:,{i+1}n{x_\i} )	\\
)	\\
\end{array}$$

In a somewhat simplified presentation,
we build the equation
$$f(\ldots, c( \:,1l{y_\i} ), \ldots) = 
g(\ldots, f(...,y_{\rho(1)},...), \ldots , f(...,y_{\rho(k)},...) ) .$$

From the i/o equations for $f$, we often\footnote{
	Our construction isn't successful in all cases.
	~
	We give a counter-example in 
	Sect.~\ref{Limitations of our approach}
}
can construct i/o equations for $g$:
~
If we have an i/o equation that matches the above equation's left-hand
side,
and we have all i/o equations needed to evaluate the recursive calls to $f$
on its right-hand side, we can build an i/o equation equation for $g$.

This way, we can reduce the problem of synthesizing a definition for
$f$ that reproduces the given i/o equations to the
problem of synthesizing a definition for $g$ from its i/o equations.
~
As a base case for this process, we may synthesize non-recursive
function definitions by anti-unification of the i/o equations.

It should be possible to prove that $f$ covers all its i/o equations
iff $g$ covers its, under some appropriate conditions.
~
We expect that a sufficient condition is that all recursive calls to
$f$ could be evaluated.
~
At least, we could demonstrate this in the above $\dup$ and $\lgth$
example.

\subsection{Termination}

In order to establish the termination of our approach, it is necessary
to define a criterion by which $g$ is easier to learn from it i/o
equations than $f$ is from its.
~
Term size or height cannot be used in a termination ordering;
when proceeding from $f$ to $g$ they may remain equal, or may even
increase, as shown in 
Fig.~\ref{Left- and right-hand term sizes of i/o equations for dup and g2} 
for the $\dup$ vs.\ $\fn{g}_2$ example.

However, the number of i/o equations decreases in this example, and in
all other ones we dealt with.
~
A sufficient criterion for this
is that $f$'s i/o equations don't all have the
same left-hand side top-most constructor.
~
However, the same criterion would have to be ensured in turn for $g$,
and it is not obvious how to achieve this.

In any case, by construction of $g$'s i/o example from $f$'s, no new
terms can arise.\footnote{
	except for the fresh left-hand side top function symbols
}
~
Even more, each term appearing in an i/o example for $g$ originates
from a right-hand side of an i/o example for $f$.
~
Therefore, our approach can't continue generating new auxiliary
functions forever, without eventually repeating the set of i/o
equations.
~
Our prototype implementation doesn't check for such repetitions,
however.

\begin{figure}
\begin{center}
$$\begin{array}{|llrr ||llrr|}
\hline
\mbox{Fct} & \mbox{Eqn} & \mbox{Lf} & \mbox{Rg} &
\mbox{Fct} & \mbox{Eqn} & \mbox{Lf} & \mbox{Rg}	\\
\hline
\hline
         & \mbox{\REF{exm dup 0}} & 2 & 1 &
\fn{g}_1 & \mbox{\REF{g1 def}}    & 1 & 1	\\
\cline{5-8}
         & \mbox{\REF{exm dup 1}} & 3 & 3 &
         & \mbox{\REF{exm g2 0}}  & 2 & 3	\\
\dup     & \mbox{\REF{exm dup 2}} & 4 & 5 &
\fn{g}_2 & \mbox{\REF{exm g2 2}}  & 4 & 5	\\
         & \mbox{\REF{exm dup 3}} & 5 & 7 &
         & \mbox{\REF{exm g2 4}}  & 6 & 7	\\
\hline
\end{array}$$
\caption{Left- and right-hand term sizes of i/o equations 
	for $\dup$ and $\fn{g}_2$}
\LABEL{Left- and right-hand term sizes of i/o equations for dup and g2}
\end{center}
\end{figure}

\section{Possible extensions}
\LABEL{Possible extensions}

In this section, we briefly sketch some possible extensions of our
approach.
~
Their investigation in detail still remains to be done.

\subsection{Extension of anti-unification}
\LABEL{Extension of anti-unification}

In Sect.~\ref{Learning functions by anti-unification}
we used syntactical anti-unification to obtain a function definition,
as a base case of our approach.
~
Several way to extend this technique can be thought of.

\paragraph*{Set anti-unification}

It can be tried to split the set of i/o equations into
disjoint subsets such that from each one an lgg
satisfying the variable condition is obtained.
~
This results in several defining equations.
~
An additional constraint might be that each subset corresponds to
another constructor symbol, observed at some given fixed position in
the left-hand side terms.

\paragraph*{Anti-unification modulo equational theory}

Another extension consists in considering an equational background
theory $E$ in anti-unification; it wasn't readily investigated in
1994.
~
See \cite{Heinz.1994a,Heinz.1994b,Heinz.1995}
for the earliest publications, and 
\cite{Burghardt.2005c,Burghardt.2017} for the latest.

As of today, the main application of $E$-anti-unification
turned out to be the synthesis of non-recursive function definitions
from input/ output equations \cite[p.3]{Burghardt.2017}.
~
To sketch an example, let $E$ consist just of definitions~\REF{add def 0},
\REF{add def s},
\REF{mul def 0},
and~\REF{mul def s}.

Assume the signature
$$\begin{array}{rrcccl}
\EQN\LABEL{sgn sq} & \sq& : & \nat & \raa & \nat        \\
\end{array}$$
and the i/o equations~\REF{exm sq 0},
\REF{exm sq 1},
\REF{exm sq 2}, and~\REF{exm sq 3}
of the squaring function.
~
Applying syntactical anti-unification to the left-hand sides yields a
variable $\x_{0123}$, and four corresponding substitutions.
~
Applying constrained $E$-generalization 
\cite[p.5, Def.2]{Burghardt.2005c}
to the right-hand sides yields a term set that contains 
$\x_{0123} \mul \x_{0123}$ as a minimal-size member,
see
Fig.~\ref{Application of $E$-anti-unification to learn squaring}.

\begin{figure}
\begin{center}
$$\begin{array}{l r@{}c@{}l c c}
\EQN\LABEL{exm sq 0}
	& \sq( & \0 & ) & = & \0	\\
\EQN\LABEL{exm sq 1}
	& \sq( & \s(\0) & ) & = & \s(\0)	\\
\EQN\LABEL{exm sq 2}
	& \sq( & \s(\s(\0)) & ) & = & \s(\s(\s(\s(\0))))	\\
\EQN\LABEL{exm sq 3}
	& \sq( & \s(\s(\s(\0))) & ) & = & \s^9(\0)	\\
\hline
\EQN & \sq( & \x_{0123} & ) & = & \x_{0123} \mul \x_{0123}	\\
\end{array}$$
\caption{Application of $E$-anti-unification to learn squaring}
\LABEL{Application of $E$-anti-unification to learn squaring}
\end{center}
\end{figure}

% choices for depth 2, 3, 4
\newcommand{\ch}[3]{%
	%\left\{%
	\begin{array}{c}%
	#1	\\[-1.0ex]%
	#2	\\[-1.0ex]%
	#3	\\%
	\end{array}%
	%\right\}%
}

\paragraph*{Depth-bounded anti-unification}

In many cases, defining equations obtained by syntactical
anti-unification appear to be too particular.
~
For example, $\s^4(\0)$ and $\s^9(\0)$ are generalized to
$\s^4(\x_{05})$, while being by $4$ greater than something wouldn't be the
first choice for a common property of both numbers for most humans.
~
As a possible remedy, a maximal depth $d$ may be introduced for the
anti-unification algorithm.
~
Beyond this depth, terms are generalized by a variable even if all their
root function symbols agree.
~
Denoting by ${\it lgg}_d(t_1,t_2)$ the result of an appropriately
modified algorithm,
it should be easy to prove that ${\it lgg}_d(t_1,t_2)$ 
can be instantiated to both $t_1$ and $t_2$, and is the most special 
term with that property among all terms of depth up to $d$.
~
If $d$ is chosen as $\infty$, ${\it lgg}_d$ and ${\it lgg}$ coincide.

\begin{figure}
\begin{center}
$$\begin{array}{|r@{}c@{}c@{}c@{}c@{}c@{}l c l@{}c@{}l@{}l|}
\hline
\sz( & \multicolumn{5}{c}{\ch{\x}{\nul}{\nul}} & )
	&=& \0 &&&	\\
\hline
\sz( & \multicolumn{5}{c}{\node(\x,\y,\z)} & )
	&=& \fn{f_1}( & \multicolumn{3}{l|}{\y,\sz(\x),\sz(\z))}	\\
\hline
\hline
\fn{f_1}( & \x &,& \ch{\y}{\0}{\0} &,& \z & ) 
	&=& \s( & \z & ) &	\\
\hline
\fn{f_1}( & \x &,& \s(\y) &,& \z & )
	&=& \fn{f_2}( & \multicolumn{3}{l|}{\x,\z,\fn{f_1}(\x,\y,\z))}	\\
\hline
\hline
\fn{f_2}( & \x &,& \ch{\y}{\0}{\0} &,& \ch{\z}{\s(\z)}{\s(\0)} & ) 
	&=& \s( & \ch{\z}{\s(\z)}{\s(\0)} & ) &	\\
\hline
\fn{f_2}( & \x &,& \ch{\y}{\s(\y)}{\s(\y)} 
	&,& \ch{\z}{\s(\z)}{\s(\s(\z))} & ) 
	&=& \s( & \ch{\z}{\s(\z)}{\s(\s(\z))} & ) &	\\
\hline
\end{array}$$
\caption{Learned tree size definition for anti-unification depth 2, 3, and 4}
\LABEL{Learned tree size definition for anti-unification depth 2, 3, and 4}
\end{center}
\end{figure}
%+++++ ORIGINAL: +++++++++++++++++++++++++++++++++++++++++++++
%
%anti_unify(ExJ,4,L=R,_Subs),
%
%FUNCTION DEFINITIONS (BEFORE SIMPLIFICATION):
%size(nl)=0
%size(nd(v10,v9,v11))=f12(v9,size(v10),size(v11))
%f12(va,0,v37)=s(v37)
%f12(v43,s(v45),v44)=f46(v43,v44,f12(v43,v45,v44))
%f46(vb,0,s(0))=s(s(0))
%f46(v63,s(v64),s(s(v65)))=s(s(s(v65)))
%
%anti_unify(ExJ,3,L=R,_Subs),
%
%FUNCTION DEFINITIONS (BEFORE SIMPLIFICATION):
%size(nl)=0
%size(nd(v10,v9,v11))=f12(v9,size(v10),size(v11))
%f12(va,0,v37)=s(v37)
%f12(v43,s(v45),v44)=f46(v43,v44,f12(v43,v45,v44))
%f46(vb,0,s(v62))=s(s(v62))
%f46(v64,s(v65),s(v66))=s(s(v66))
%
%anti_unify(ExJ,2,L=R,_Subs),
%
%FUNCTION DEFINITIONS (BEFORE SIMPLIFICATION):
%size(v2)=0
%size(nd(v9,v8,v10))=f11(v8,size(v9),size(v10))
%f11(v36,v37,v38)=s(v38)
%f11(v44,s(v46),v45)=f47(v44,v45,f11(v44,v46,v45))
%f47(v63,v64,v65)=s(v65)
%f47(v67,v68,v69)=s(v69)
%
%8 UNCOVERED I/O EXAMPLES
%
%anti_unify(ExJ,1,L=R,_Subs),
%
%FAILED

In our prototype implementation, we meanwhile built in such a depth
boundary.
~
Figure~\ref{Learned tree size definition for anti-unification depth 2, 3, and 4}
compares the learned function definitions for $\sz$
for $d=2,3,4$ (top to bottom).
~
For example, for $d=2$, the ---nonsensical--- equation $\sz(\x)=\0$ is
learned, while for $d \geq 3$ the respective equation reads
$\sz(\nul)=\0$.
~
Not surprisingly, for $d=2$ only one of the given $9$ i/o equations is
covered.
~
For $d \leq 1$, the attempt to learn defining equations for $\sz$
fails.

For $d = 4$, the learned equations agree with those for $d = \infty$,
and hence also with those for all intermediate depths.
~
The prototype run for $d = \infty$ is shown in App.~\ref{Size of a tree}.
~
Note that the prototype simplifies equations by removing irrelevant
function arguments.
~
For this reason, {\tt f12} there has only two arguments, while the
corresponding function $\fn{f_1}$ in
Fig.~\ref{Learned tree size definition for anti-unification depth 2, 3, and 4}
has three.

\subsection{Extension of structural recursion}
\LABEL{Extension of structural recursion}

Some functions are best defined by simultaneous recursion on several
arguments.
~
As an example, consider the sort definition~\REF{sort blist}
with $\Bnil$, $\Bo$, and $\Bi$ denoting an empty list, a $0$ digit,
and a $1$ digit, respectively.
~
For technical reasons, such a list is interpreted in reversed order,
e.g.\ $\Bo(\Bi(\Bi(\Bnil)))$ denotes the number $6$.
~
The sum function $\Badd$, its signature shown in~\REF{sgn badd},
may then be defined by the following rewrite system:
$$\begin{array}{r r@{}c@{}c@{}c@{}c c l}
\EQN & \Badd( & \x & , & \Bnil & ) & = & \x	\\
\EQN & \Badd( & \Bnil & , & \y & ) & = & \y	\\
\EQN & \Badd( & \Bo(\x) & , & \Bo(\y) & ) & = & \Bo(\Badd(\x,\y))	\\
\EQN & \Badd( & \Bo(\x) & , & \Bi(\y) & ) & = & \Bi(\Badd(\x,\y))	\\
\EQN & \Badd( & \Bi(\x) & , & \Bo(\y) & ) & = & \Bi(\Badd(\x,\y))	\\
\EQN & \Badd( & \Bi(\x) & , & \Bi(\y) & ) & = & \Bo(\Binc(\Badd(\x,\y))) \\
\end{array}$$
where
$$\begin{array}{rrcccl}
\EQN
	& \Binc& : & \Blist& \raa & \Blist	\\
\end{array}$$
is a function to increment a binary digit list.
~
This corresponds to the usual hardware implementation, with $\Binc$
being used for the carry.

It is obvious that this definition cannot be obtained from our
simple structural recursion scheme from 
Sect.~\ref{Learning functions by structural recursion},
neither by recurring over argument position $1$ nor over $2$.
~
Instead, we would need recursion over both positions simultaneously,
i.e.\ a scheme like
$$\begin{array}{r r@{}c@{}c@{}c@{}c c l}
\EQN\LABEL{rec badd 1} & \Badd(&\Bnil&,&\Bnil&) & = & \fn{g_{15}} \\
\EQN\LABEL{rec badd 2} & \Badd(&\Bnil&,&\Bo(\y)&) & = & \fn{g_{16}(\y)} \\
\EQN\LABEL{rec badd 3} & \Badd(&\Bnil&,&\Bi(\y)&) & = & \fn{g_{17}(\y)} \\
\EQN\LABEL{rec badd 4} & \Badd(&\Bo(\x)&,&\Bnil&) & = & \fn{g_{18}(\x)} \\
\EQN\LABEL{rec badd 5} 
	& \Badd(&\Bo(\x)&,&\Bo(\y)&) & = & \fn{g_{19}}(\Badd(\x,\y)) \\
\EQN\LABEL{rec badd 6} 
	& \Badd(&\Bo(\x)&,&\Bi(\y)&) & = & \fn{g_{20}}(\Badd(\x,\y)) \\
\EQN\LABEL{rec badd 7} & \Badd(&\Bi(\x)&,&\Bnil&) & = & \fn{g_{21}(\x)} \\
\EQN\LABEL{rec badd 8} 
	& \Badd(&\Bi(\x)&,&\Bo(\y)&) & = & \fn{g_{22}}(\Badd(\x,\y)) \\
\EQN\LABEL{rec badd 9} 
	& \Badd(&\Bi(\x)&,&\Bi(\y)&) & = & \fn{g_{23}}(\Badd(\x,\y)) \\
\end{array}$$

An extension of our approach could provide such a scheme, additionally
to the simple structural recursion scheme.

If we could prove that each function definition obtainable by the 
simple recursion scheme can also be obtained by a simultaneous
recursion scheme, we needed only to employ the latter.
~
This way, we would no longer need to guess an appropriate argument
position to recur over; instead we could always recur simultaneously over
all arguments of a given sort.
~
Unfortunately, simultaneous recursion is not stronger than simple
structural recursion.
~
For example, the function $\app$ to concatenate two given lists can be
obtained by simple recursion over the first argument
(see~\REF{app def nil},\REF{app def cons} 
in Fig.~\ref{Example function definitions}),
but not by simultaneous recursion:
$\app(\w \cons \x, \y \cons \z) = \fn{g_{24}}(\w,\y,\app(\x,\z))$
doesn't lead to a sensible definition, for any choice of $\fn{g_{24}}$.

One possible remedy is to try simple structural recursion first, on
any appropriate argument position, and simultaneous recursion next, on
any appropriate set of argument positions.
~
Alternatively, user commands may be required about which recursion to
try on which argument position(s).

\newcommand{\undef}{\Omega}

Another possibility might be to employ a fully general structural
recursion scheme, like
$$\begin{array}{rr rc rcccccc l}
& \EQN
	& \app(\w \cons \x, \y \cons \z)
	& =
	& \fn{g_{24}}(\w,\y,
	& \app(\w \cons \x,\z)
	& , 
	& \app(\x,\y \cons \z)
	& ,
	& \app(\x,\z)
	& )	\\
\mbox{and }
& \EQN
	& \Badd(\Bo(\x), \Bo(\y))
	& =
	& \fn{g_{25}}(
	& \Badd(\Bo(\x),\y)
	& , 
	& \Badd(\x,\Bo(\y))
	& ,
	& \Badd(\x,\y)
	& )
	& .	\\
\end{array}$$
In this scheme, calls for simple recursion over each
position are provided, as well as for simultaneous recursion over each
position set.
~
A new symbol $\undef$, intended to denote an undefined term,
could be added to the term language.
~
When e.g.\ i/o equations are missing to compute $\Badd(\Bo(\x),\y)$
for some particular instance, the first argument of $\fn{g_{25}}$
would be set to $\undef$ in the respective i/o equation.
~
In syntactical anti-unification and coverage test,
$\undef$ needed to be handled appropriately.
~
This way, only one recursion scheme would be needed, and no choice
of appropriate argument position(s) would be necessary.
~
However, arities of auxiliary functions might grow exponentially.

\subsection{Limitations of our approach}
\LABEL{Limitations of our approach}

In this section, we demonstrate an example where our approach fails.
~
Consider again the squaring function, its signature shown in~\REF{sgn sq},
and consider again its i/o equations~\REF{exm sq 0},
\REF{exm sq 1},
\REF{exm sq 2}, and~\REF{exm sq 3}.

Since syntactical 
anti-unification as in 
Sect.~\ref{Learning functions by anti-unification}
(i.e.\ not considering an equational background theory $E$)
doesn't lead to a valid function definition, we
build a structural recursion scheme as in
Sect.~\ref{Learning functions by structural recursion}:
$$\begin{array}{rrcl}
\EQN\LABEL{rec sq 1} & \sq(\0) & = & \fn{g_{11}} \\
\EQN\LABEL{rec sq 2} & \sq(\s(\x)) & = & \fn{g_{12}}(\sq(\x))   \\
\end{array}$$
We get $\fn{g_{11}} = \0$,
and the following i/o equations for $\fn{g_{12}}$:
$$\begin{array}{r ccccccc}
\EQN\LABEL{exm g12 0}
        & \s(\0) 
        & \stackrel{\REF{exm sq 1}}{=} & \sq( \s(\0) )
        & \stackrel{\REF{rec sq 2}}{=} & \fn{g_{12}}(\sq(\0))
        & \stackrel{\REF{exm sq 0}}{=} & \fn{g_{12}}(\0)  \\
\EQN\LABEL{exm g12 1}
        & \s(\s(\s(\s(\0)))) 
        & \stackrel{\REF{exm sq 2}}{=} & \sq( \s(\s(\0)) )
        & \stackrel{\REF{rec sq 2}}{=} & \fn{g_{12}}(\sq(\s(\0)))
        & \stackrel{\REF{exm sq 1}}{=} & \fn{g_{12}}(\s(\0))   \\
\EQN\LABEL{exm g12 4}
        & \s^9(\0)
        & \stackrel{\REF{exm sq 3}}{=} & \sq( \s(\s(\s(\0))) )
        & \stackrel{\REF{rec sq 2}}{=} & \fn{g_{12}}(\sq(\s(\s(\0))))
        & \stackrel{\REF{exm sq 2}}{=} & \fn{g_{12}}(\s(\s(\s(\s(\0))))) \\
\end{array}$$
Observe that we are able to obtain i/o equations for $\fn{g_{12}}$
only on square numbers.
~
For example, there is no obvious way to
determine the value of $\fn{g_{12}}(\s(\s(\s(\0))))$.

Syntactically 
anti-unifying $\fn{g_{12}}$'s i/o equation still doesn't yield a valid
function definition.
~
So we set up a recursion scheme for $\fn{g_{12}}$, in turn:
$$\begin{array}{rrcl}
\EQN\LABEL{rec g12 1} & \fn{g_{12}}(\0) & = & \fn{g_{13}} \\
\EQN\LABEL{rec g12 2} 
	& \fn{g_{12}}(\s(\x)) & = & \fn{g_{14}}(\fn{g_{12}}(\x))   \\
\end{array}$$
Again, $\fn{g_{13}} = \s(\0)$ is obvious.
~
Trying to obtain i/o equations for $\fn{g_{14}}$,
we get stuck, since we don't know how $\fn{g_{12}}$ should behave on
non-square numbers:
$$\begin{array}{r ccccccc}
\EQN\LABEL{exm g14 1}
        & \s(\s(\s(\s(\0)))) 
        & \stackrel{\REF{exm g12 1}}{=} & \fn{g_{12}}( \s(\0) )
        & \stackrel{\REF{rec g12 2}}{=} & \fn{g_{14}}(\fn{g_{12}}(\0))
        & \stackrel{\REF{exm g12 0}}{=} & \fn{g_{14}}(\s(\0))  \\
\EQN\LABEL{exm g14 4}
        & ??
        & \stackrel{??}{=} & \fn{g_{12}}( \s(\s(\0)) )
        & \stackrel{\REF{rec g12 2}}{=} & \fn{g_{14}}(\fn{g_{12}}(\s(\0)))
        & \stackrel{\REF{exm g12 1}}{=} 
	& \fn{g_{14}}(\s(\s(\s(\s(\0)))))   \\
\EQN\LABEL{exm g14 ?2}
        & ??
        & \stackrel{??}{=}
	& \fn{g_{12}}( \s(\s(\s(\0))) )
        & \stackrel{\REF{rec g12 2}}{=}
	& \fn{g_{14}}(\fn{g_{12}}(\s(\s(\0))))
        & \stackrel{??}{=}
	& \fn{g_{14}}(??) \\
\EQN\LABEL{exm g14 ?3}
        & \s^9(\0)
        & \stackrel{\REF{exm g12 4}}{=}
	& \fn{g_{12}}( \s(\s(\s(\s(\0)))) )
        & \stackrel{\REF{rec g12 2}}{=}
	& \fn{g_{14}}(\fn{g_{12}}(\s(\s(\s(\0)))))
        & \stackrel{??}{=}
	& \fn{g_{14}}(??) \\
\end{array}$$
As an alternative,
by applying~\REF{rec g12 2} sufficiently often rather than just once,
we can obtain:
$$\begin{array}{r cccccccccccc}
\EQN\LABEL{exm g14 pow}
        & \s^9(\0)
        & \stackrel{\REF{exm g12 4}}{=}
	& \fn{g_{12}}( \s(\s(\s(\s(\0)))) )	\\
        && \stackrel{\REF{rec g12 2}}{=}
	& \fn{g_{14}}(\fn{g_{12}}(\s(\s(\s(\0)))))	\\
        && \stackrel{\REF{rec g12 2}}{=}
	& \fn{g_{14}}(\fn{g_{14}}(\fn{g_{12}}(\s(\s(\0)))))	\\
        && \stackrel{\REF{rec g12 2}}{=}
	& \fn{g_{14}}(\fn{g_{14}}(\fn{g_{14}}(\fn{g_{12}}(\s(\0)))))
        & \stackrel{\REF{exm g12 1}}{=}
	& \fn{g_{14}}(\fn{g_{14}}(\fn{g_{14}}(\s(\s(\s(\s(\0))))))) \\
\end{array}$$
However, no approach is known to learn $\fn{g_{14}}$ from an extended
i/o equation like~\REF{exm g14 pow}, which determines 
$\fn{g_{14}} \circ \fn{g_{14}} \circ \fn{g_{14}}$
rather than $\fn{g_{14}}$ itself.
~
In such cases, we resort to the excuse that the original function,
$\sq$ isn't definable by structural recursion.

A precise criterion for the class that our approach can handle is
still to be found.
~
It is not even clear that such a criterion can be computable.
~
If not, it should still be possible to give computable necessary and
sufficient approximations.

% \section{Other Approaches}
% 
% {\sc Igor 2}
% \cite{Kitzelmann.2010}
% %,Hofmann.2010b,Hofmann.Kitzelmann.2010,Kitzelmann.2008a,Kitzelmann.2008b
% refines the result of syntactical anti-unification of the i/o
% equations (or an appropriate subset thereof)
% by instantiating unbound right-hand side variables step by step with known
% background functions or fresh auxiliary functions to be synthesized
% lateron.
% ~
% As a consequence, 
% constructors common to all i/o equation right-hand sides will appear in the
% learned rules at the same positions.
% 
% %\cite{Barzdins.1991}
% %\cite{Biermann.1978}
% \cite{Biermann.1985}	% overview in ch.4
% %\cite{Biermann.Krishnaswamy.1976}
% %\cite{Biermann.Petry.1975}
% \cite{Edelkamp.Schrodl.1999}	% learning from traces via DFA
% %\cite{Feng.Muggleton.1990}	% golem
% \cite{Guiho.Jouannaud.Treuil.1977}	% sisp1, minimal exm cnt
% \cite{Hirschberger.Hofmann.Kitzelmann.2007}	% comparison adate, atre, dialogs2
% \cite{Hofmann.2010b}	% igor2
% \cite{Hofmann.Kitzelmann.Schmid.2009b}	% igor2 ?
% \cite{Kitzelmann.Schmid.2006b}
% \cite{Popelinsky.1994}
% \cite{Schmid.2001}
% \cite{Schmid.Wysotzki.1998}
% \cite{Summers.1977}

\newpage

\appendix

\section{Example runs of our prototype implementation}
\LABEL{Example runs of our prototype implementation}

\subsection{Addition of $0$-$s$ numbers}
\LABEL{Addition of $0$-$s$ numbers}

\tiny

\begin{verbatim}

?- SgI = [ + signature [nat,nat] --> nat],
|    SD  = [ nat sortdef 0 ! s(nat)],
|    ExI = [ 0       + 0             = 0,
|            s(0)    + 0             = s(0),
|            0       + s(0)          = s(0),
|            0       + s(s(0))       = s(s(0)),
|            s(0)    + s(0)          = s(s(0)),
|            s(0)    + s(s(0))       = s(s(s(0))),
|            s(s(0)) + s(0)          = s(s(s(0))),
|            s(s(0)) + 0             = s(s(0))],
|    run(+,SgI,SD,ExI).
+++++ Examples input check:
+++++ Example 1:
+++++ Example 2:
+++++ Example 3:
+++++ Example 4:
+++++ Example 5:
+++++ Example 6:
+++++ Example 7:
+++++ Example 8:
+++++ Examples input check done
induce(+)
. trying argument position:     1
. inducePos(+,1,0)
. . matching examples:  [0+0=0,0+s(0)=s(0),0+s(s(0))=s(s(0))]
. . anti-unifier:       0+v3 = v3
. inducePos(+,1,0)
. inducePos(+,1,s(nat))
. . matching examples:  [s(0)+0=s(0),s(0)+s(0)=s(s(0)),s(0)+s(s(0))=s(s(s(0))),s(s(0))+s(0)=s(s(s(0))),s(s(0))+0=s(s(0))]
. . new recursion scheme:       s(v9)+v8 = f10(v8,v9+v8)
. . derive new equation:        s(0) = s(0)+0 = f10(0,0)
. . derive new equation:        s(s(0)) = s(0)+s(0) = f10(s(0),s(0))
. . derive new equation:        s(s(s(0))) = s(0)+s(s(0)) = f10(s(s(0)),s(s(0)))
. . derive new equation:        s(s(s(0))) = s(s(0))+s(0) = f10(s(0),s(s(0)))
. . derive new equation:        s(s(0)) = s(s(0))+0 = f10(0,s(0))
. . induce(f10)
. . . trying argument position: 1
. . . inducePos(f10,1,0)
. . . . matching examples:      [f10(0,0)=s(0),f10(0,s(0))=s(s(0))]
. . . . anti-unifier:   f10(0,v13) = s(v13)
. . . inducePos(f10,1,0)
. . . inducePos(f10,1,s(nat))
. . . . matching examples:      [f10(s(0),s(0))=s(s(0)),f10(s(0),s(s(0)))=s(s(s(0))),f10(s(s(0)),s(s(0)))=s(s(s(0)))]
. . . . anti-unifier:   f10(s(v15),s(v16)) = s(s(v16))
. . . inducePos(f10,1,s(nat))
. . . all examples covered
. . induce(f10)
. inducePos(+,1,s(nat))
. all examples covered
induce(+)
+++++ Examples output check:
+++++ Examples output check done
FUNCTION SIGNATURES:
f10 signature [nat,nat]-->nat
(+)signature[nat,nat]-->nat

FUNCTION EXAMPLES:
0+0=0
s(0)+0=s(0)
0+s(0)=s(0)
0+s(s(0))=s(s(0))
s(0)+s(0)=s(s(0))
s(0)+s(s(0))=s(s(s(0)))
s(s(0))+s(0)=s(s(s(0)))
s(s(0))+0=s(s(0))

FUNCTION DEFINITIONS:
0+v17=v17
s(v18)+v19=f10(v19,v18+v19)
f10(0,v20)=s(v20)
f10(s(v21),s(v22))=s(s(v22))

?- 

\end{verbatim}

\subsection{Size of a tree}
\LABEL{Size of a tree}

\tiny

\begin{verbatim}


?- SgI = [ size signature [tree] --> nat],
|    SD  = [ tree sortdef nl ! nd(tree,nat,tree),
|            nat  sortdef 0 ! s(nat)],
|    ExI = [ size(nl)                                = 0,
|            size(nd(nl,va,nl))                      = s(0),
|            size(nd(nd(nl,va,nl),vb,nl))    = s(s(0)),
|            size(nd(nl,va,nd(nl,vb,nl)))    = s(s(0)),
|            size(nd(nd(nl,va,nl),vb,nd(nl,vc,nl)))  = s(s(s(0))),
|            size(nd(nl,va,nd(nd(nl,vb,nl),vc,nl)))  = s(s(s(0))),
|            size(nd(nl,va,nd(nl,vb,nd(nl,vc,nl))))  = s(s(s(0))),
|            size(nd(nd(nl,va,nl),vb,nd(nd(nl,vc,nl),vd,nl)))        = s(s(s(s(0)))),
|            size(nd(nd(nd(nl,va,nl),vb,nl),vc,nd(nl,vd,nl)))        = s(s(s(s(0))))
|            ],
|    run(size,SgI,SD,ExI).
+++++ Examples input check:
+++++ Example 1:
+++++ Example 2:
+++++ Example 3:
+++++ Example 4:
+++++ Example 5:
+++++ Example 6:
+++++ Example 7:
+++++ Example 8:
+++++ Example 9:
Variable sorts:
[vd:nat,vc:nat,vb:nat,va:nat]
+++++ Examples input check done
induce(size)
. trying argument position:     1
. inducePos(size,1,nl)
. . matching examples:  [size(nl)=0]
. . anti-unifier:       size(nl) = 0
. inducePos(size,1,nl)
. inducePos(size,1,nd(tree,nat,tree))
. . matching examples:  [size(nd(nl,va,nl))=s(0),size(nd(nd(nl,va,nl),vb,nl))=s(s(0)),size(nd(nl,va,nd(nl,vb,nl)))=s(s(0)),size(nd(nd(nl,va,nl),vb,n...
. . new recursion scheme:       size(nd(v10,v9,v11)) = f12(v9,size(v10),size(v11))
. . derive new equation:        s(0) = size(nd(nl,va,nl)) = f12(va,0,0)
. . derive new equation:        s(s(0)) = size(nd(nd(nl,va,nl),vb,nl)) = f12(vb,s(0),0)
. . derive new equation:        s(s(0)) = size(nd(nl,va,nd(nl,vb,nl))) = f12(va,0,s(0))
. . derive new equation:        s(s(s(0))) = size(nd(nd(nl,va,nl),vb,nd(nl,vc,nl))) = f12(vb,s(0),s(0))
. . derive new equation:        s(s(s(0))) = size(nd(nl,va,nd(nd(nl,vb,nl),vc,nl))) = f12(va,0,s(s(0)))
. . derive new equation:        s(s(s(0))) = size(nd(nl,va,nd(nl,vb,nd(nl,vc,nl)))) = f12(va,0,s(s(0)))
. . derive new equation:        s(s(s(s(0)))) = size(nd(nd(nl,va,nl),vb,nd(nd(nl,vc,nl),vd,nl))) = f12(vb,s(0),s(s(0)))
. . derive new equation:        s(s(s(s(0)))) = size(nd(nd(nd(nl,va,nl),vb,nl),vc,nd(nl,vd,nl))) = f12(vc,s(s(0)),s(0))
. . induce(f12)
. . . trying argument position: 1
. . . inducePos(f12,1,0)
. . . . matching examples:      []
. . . . no examples
. . . inducePos(f12,1,0)
. . . inducePos(f12,1,s(nat))
. . . . matching examples:      []
. . . . no examples
. . . inducePos(f12,1,s(nat))
. . . uncovered examples:       [f12(va,0,0)=s(0),f12(va,0,s(0))=s(s(0)),f12(va,0,s(s(0)))=s(s(s(0))),f12(vb,s(0),0)=s(s(0)),f12(vb,s(0),s(0))=s(s(s...
. . . trying argument position: 2
. . . inducePos(f12,2,0)
. . . . matching examples:      [f12(va,0,0)=s(0),f12(va,0,s(0))=s(s(0)),f12(va,0,s(s(0)))=s(s(s(0)))]
. . . . anti-unifier:   f12(va,0,v37) = s(v37)
. . . inducePos(f12,2,0)
. . . inducePos(f12,2,s(nat))
. . . . matching examples:      [f12(vb,s(0),0)=s(s(0)),f12(vb,s(0),s(0))=s(s(s(0))),f12(vb,s(0),s(s(0)))=s(s(s(s(0)))),f12(vc,s(s(0)),s(0))=s(s(s(s...
. . . . new recursion scheme:   f12(v43,s(v45),v44) = f46(v43,v44,f12(v43,v45,v44))
. . . . derive new equation:    s(s(0)) = f12(vb,s(0),0) = f46(vb,0,s(0))
. . . . derive new equation:    s(s(s(0))) = f12(vb,s(0),s(0)) = f46(vb,s(0),s(s(0)))
. . . . derive new equation:    s(s(s(s(0)))) = f12(vb,s(0),s(s(0))) = f46(vb,s(s(0)),s(s(s(0))))
. . . . derive new equation:    s(s(s(s(0)))) = f12(vc,s(s(0)),s(0)) = f46(vc,s(0),s(s(s(0))))
. . . . induce(f46)
. . . . . trying argument position:     1
. . . . . inducePos(f46,1,0)
. . . . . . matching examples:  []
. . . . . . no examples
. . . . . inducePos(f46,1,0)
. . . . . inducePos(f46,1,s(nat))
. . . . . . matching examples:  []
. . . . . . no examples
. . . . . inducePos(f46,1,s(nat))
. . . . . uncovered examples:   [f46(vb,0,s(0))=s(s(0)),f46(vb,s(0),s(s(0)))=s(s(s(0))),f46(vb,s(s(0)),s(s(s(0))))=s(s(s(s(0)))),f46(vc,s(0),s(s(s(0...
. . . . . trying argument position:     2
. . . . . inducePos(f46,2,0)
. . . . . . matching examples:  [f46(vb,0,s(0))=s(s(0))]
. . . . . . anti-unifier:       f46(vb,0,s(0)) = s(s(0))
. . . . . inducePos(f46,2,0)
. . . . . inducePos(f46,2,s(nat))
. . . . . . matching examples:  [f46(vb,s(0),s(s(0)))=s(s(s(0))),f46(vb,s(s(0)),s(s(s(0))))=s(s(s(s(0)))),f46(vc,s(0),s(s(s(0))))=s(s(s(s(0))))]
. . . . . . anti-unifier:       f46(v63,s(v64),s(s(v65))) = s(s(s(v65)))
. . . . . inducePos(f46,2,s(nat))
. . . . . all examples covered
. . . . induce(f46)
. . . inducePos(f12,2,s(nat))
. . . all examples covered
. . induce(f12)
. inducePos(size,1,nd(tree,nat,tree))
. all examples covered
induce(size)
+++++ Examples output check:
+++++ Examples output check done
FUNCTION SIGNATURES:
f46 signature [nat,nat,nat]-->nat
f12 signature [nat,nat,nat]-->nat
size signature [tree]-->nat

FUNCTION EXAMPLES:
size(nl)=0
size(nd(nl,va,nl))=s(0)
size(nd(nd(nl,va,nl),vb,nl))=s(s(0))
size(nd(nl,va,nd(nl,vb,nl)))=s(s(0))
size(nd(nd(nl,va,nl),vb,nd(nl,vc,nl)))=s(s(s(0)))
size(nd(nl,va,nd(nd(nl,vb,nl),vc,nl)))=s(s(s(0)))
size(nd(nl,va,nd(nl,vb,nd(nl,vc,nl))))=s(s(s(0)))
size(nd(nd(nl,va,nl),vb,nd(nd(nl,vc,nl),vd,nl)))=s(s(s(s(0))))
size(nd(nd(nd(nl,va,nl),vb,nl),vc,nd(nl,vd,nl)))=s(s(s(s(0))))

FUNCTION DEFINITIONS:
size(nl)=0
size(nd(v66,v67,v68))=f12(size(v66),size(v68))
f12(0,v69)=s(v69)
f12(s(v70),v71)=f46(v71,f12(v70,v71))
f46(0,s(0))=s(s(0))
f46(s(v72),s(s(v73)))=s(s(s(v73)))

?- 

\end{verbatim}

% truncation ruler:
%23456789 123456789 123456789 123456789 123456789 123456789 123456789 123456789 123456789 123456789 123456789 123456789 123456789 123456789 12345678...
% truncated closing parantheses: ))))))))))) ]]]]

\subsection{Reversing a list}
\LABEL{Reversing a list}

\tiny

\begin{verbatim}

?- SgI = [rev signature [list] --> list],
|    SD  = [ list sortdef [] ! [nat|list],
|            nat  sortdef  0 ! s(nat)],
|    ExI = [ rev([])         = [], 
|            rev([va])       = [va],
|            rev([vb,va])    = [va,vb],
|            rev([vc,vb,va]) = [va,vb,vc]],
|    run(rev,SgI,SD,ExI).
+++++ Examples input check:
+++++ Example 1:
+++++ Example 2:
+++++ Example 3:
+++++ Example 4:
Variable sorts:
[vc:nat,vb:nat,va:nat]
+++++ Examples input check done
induce(rev)
. trying argument position:     1
. inducePos(rev,1,[])
. . matching examples:  [rev([])=[]]
. . anti-unifier:       rev([]) = []
. inducePos(rev,1,[])
. inducePos(rev,1,[nat|list])
. . matching examples:  [rev([va])=[va],rev([vb,va])=[va,vb],rev([vc,vb,va])=[va,vb,vc]]
. . new recursion scheme:       rev([v7|v8]) = f9(v7,rev(v8))
. . derive new equation:        [va] = rev([va]) = f9(va,[])
. . derive new equation:        [va,vb] = rev([vb,va]) = f9(vb,[va])
. . derive new equation:        [va,vb,vc] = rev([vc,vb,va]) = f9(vc,[va,vb])
. . induce(f9)
. . . trying argument position: 1
. . . inducePos(f9,1,0)
. . . . matching examples:      []
. . . . no examples
. . . inducePos(f9,1,0)
. . . inducePos(f9,1,s(nat))
. . . . matching examples:      []
. . . . no examples
. . . inducePos(f9,1,s(nat))
. . . uncovered examples:       [f9(va,[])=[va],f9(vb,[va])=[va,vb],f9(vc,[va,vb])=[va,vb,vc]]
. . . trying argument position: 2
. . . inducePos(f9,2,[])
. . . . matching examples:      [f9(va,[])=[va]]
. . . . anti-unifier:   f9(va,[]) = [va]
. . . inducePos(f9,2,[])
. . . inducePos(f9,2,[nat|list])
. . . . matching examples:      [f9(vb,[va])=[va,vb],f9(vc,[va,vb])=[va,vb,vc]]
. . . . new recursion scheme:   f9(v22,[v23|v24]) = f25(v22,v23,f9(v22,v24))
. . . . derive new equation:    [va,vb] = f9(vb,[va]) = f25(vb,va,[vb])
. . . . derive new equation:    [va,vb,vc] = f9(vc,[va,vb]) = f25(vc,va,[vb,vc])
. . . . induce(f25)
. . . . . trying argument position:     1
. . . . . inducePos(f25,1,0)
. . . . . . matching examples:  []
. . . . . . no examples
. . . . . inducePos(f25,1,0)
. . . . . inducePos(f25,1,s(nat))
. . . . . . matching examples:  []
. . . . . . no examples
. . . . . inducePos(f25,1,s(nat))
. . . . . uncovered examples:   [f25(vb,va,[vb])=[va,vb],f25(vc,va,[vb,vc])=[va,vb,vc]]
. . . . . trying argument position:     2
. . . . . inducePos(f25,2,0)
. . . . . . matching examples:  []
. . . . . . no examples
. . . . . inducePos(f25,2,0)
. . . . . inducePos(f25,2,s(nat))
. . . . . . matching examples:  []
. . . . . . no examples
. . . . . inducePos(f25,2,s(nat))
. . . . . uncovered examples:   [f25(vb,va,[vb])=[va,vb],f25(vc,va,[vb,vc])=[va,vb,vc]]
. . . . . trying argument position:     3
. . . . . inducePos(f25,3,[])
. . . . . . matching examples:  []
. . . . . . no examples
. . . . . inducePos(f25,3,[])
. . . . . inducePos(f25,3,[nat|list])
. . . . . . matching examples:  [f25(vb,va,[vb])=[va,vb],f25(vc,va,[vb,vc])=[va,vb,vc]]
. . . . . . anti-unifier:       f25(v37,va,[vb|v38]) = [va,vb|v38]
. . . . . inducePos(f25,3,[nat|list])
. . . . . all examples covered
. . . . induce(f25)
. . . inducePos(f9,2,[nat|list])
. . . all examples covered
. . induce(f9)
. inducePos(rev,1,[nat|list])
. all examples covered
induce(rev)
+++++ Examples output check:
+++++ Examples output check done
FUNCTION SIGNATURES:
f25 signature [nat,nat,list]-->list
f9 signature [nat,list]-->list
rev signature [list]-->list

FUNCTION EXAMPLES:
rev([])=[]
rev([va])=[va]
rev([vb,va])=[va,vb]
rev([vc,vb,va])=[va,vb,vc]

FUNCTION DEFINITIONS:
rev([])=[]
rev([v39|v40])=f9(v39,rev(v40))
f9(v41,[])=[v41]
f9(v42,[v43|v44])=f25(v43,f9(v42,v44))
f25(v41,[v45|v46])=[v41,v45|v46]

?-

\end{verbatim}

\normalsize

\newpage

\bibliographystyle{alpha}
\bibliography{lit}

\end{document}